\documentclass[a4paper,twocolumn,
english,aps,pre,floatfix,superscriptaddress,showpacs,nofootinbib]{revtex4-1}
\usepackage[T1]{fontenc}
\usepackage[latin1]{inputenc}
\usepackage{amsmath}
\usepackage{babel}
\usepackage{graphics}
\usepackage{amssymb}


\begin{document}
\title
{Smoothly-varying hopping rates in driven flow with exclusion  
}
\author {R. B. \surname{Stinchcombe}}
\email{r.stinchcombe1@physics.ox.ac.uk}
\affiliation{Rudolf Peierls Centre for Theoretical Physics, University of
Oxford, 1 Keble Road, Oxford OX1 3NP, United Kingdom}
\author {S. L. A. \surname{de Queiroz}}
\email{sldq@if.ufrj.br}
\affiliation{Rudolf Peierls Centre for Theoretical Physics, University of
Oxford, 1 Keble Road, Oxford OX1 3NP, United Kingdom}
\affiliation{Instituto de F\'\i sica, Universidade Federal do
Rio de Janeiro, Caixa Postal 68528, 21941-972
Rio de Janeiro RJ, Brazil}

\date{\today}

\begin{abstract}
We consider the one-dimensional totally asymmetric simple exclusion process (TASEP)
with position-dependent hopping rates. The problem is solved,  in a
 mean field/adiabatic approximation, for a general (smooth) form of spatial rate 
variation. Numerical simulations of systems with hopping rates varying linearly
against position (constant rate gradient), for both periodic and open boundary
conditions, provide detailed confirmation of theoretical predictions, concerning 
steady-state average density profiles and currents, as well as open-system phase 
boundaries, to excellent numerical accuracy.
\end{abstract}
\maketitle
 
\section{Introduction} 
\label{intro}
In this paper we investigate the one-dimensional totally asymmetric
simple exclusion process (TASEP)~\cite{derr98}, in the presence of non-uniform
hopping rates. The TASEP is a biased diffusion process
for particles with hard-core repulsion 
(excluded volume)~\cite{derr98,sch00,derr93,be07}.
Notwithstanding the simplicity of formulation of its basic rules, this model
can exhibit a wealth of non-trivial properties, and is 
considered a paradigm in the field of non-equilibrium phenomena. 
Quenched random inhomogeneities in the TASEP have been extensively considered 
earlier~\cite{tb98,bbzm99,k00,ssl04,ed04,hs04,dqrbs08,gs08}. In contrast,
the case of deterministically-varying, position-dependent physical parameters 
has received  less attention~\cite{lak05,lak06}.

The TASEP and its generalizations have been
applied to a broad range of non-equilibrium physical contexts, 
from macroscopic ones such as highway traffic~\cite{sz95} to microscopic ones,
including sequence alignment in computational biology~\cite{rb02}
and current shot noise in quantum-dot chains~\cite{kvo10}. Situations 
may arise where monotonic spatial variations 
in an associated parameter can be relevant (such as gradients in the first case, and
the "gap-cost", or an applied electric field, for the latter two cases).
By contrast, the effects of, e.g., temperature
gradients on the equilibrium~\cite{pkt07} and transport~\cite{grant}
properties of spin systems have been studied in detail; the same applies
to concentration gradients in percolation~\cite{sapoval,nolin08,gastner10}. 
One typically gets a picture of spatial phase separation, in which a 
high-temperature (or low-concentration) disordered region connects to a 
low-temperature (high-concentration) ordered one 
via an interface, whose features (e.g., width)  
scale in a non-trivial way with the inhomogeneity parameters.
More recently, experimental progress in cold-atom trapping~\cite{trapexp} has been
one motivation behind  the theoretical study of (pseudo)--spin systems in 
trapping potentials such as magnetic fields with a wedge-like or parabolic 
profile~\cite{pbs08,cv09,trap10}.  

We consider the problem of flow with exclusion, for which the time  evolution 
of the $1+1$ dimensional TASEP is the fundamental discrete model.
The particle number $n_\ell$ at lattice site $\ell$ can be $0$ or $1$, 
and the forward hopping of particles is only to an empty adjacent site. 
The current across the bond from $\ell$ to $\ell +1$ depends also
on the stochastic attempt rate, $p_\ell$, associated to it and is thus
given by  $J_{\ell,\ell+1}= p_\ell\,n_\ell (1-n_{\ell+1})$~. 
For the usual homogeneous case of $p_\ell =p$, in numerical simulations 
one can effectively make $p=1$, provided that the inherent stochasticity of 
the process is kept, via e.g. random selection of site occupation 
update~\cite{dqrbs08}. This amounts to a trivial renormalization 
of the time scale. 

Here, we consider a position-dependent hopping
rate (which cannot thus be simply renormalized away).
By using periodic or open boundary conditions, with assorted overall
densities in the former case, and injection/ejection rates in the latter,
we investigate the consequent effects upon the associated particle 
density profiles and currents.

To begin with we give the generic dynamic mean field theory for arbitrary "slow" 
space-dependence of the hopping rate. 
We then turn, for more specific results, to the steady state
in the case of a linear dependence of 
$p_\ell$ on position (uniform gradient).
It is remarkable that, from the combination of the mean field approach with an
adiabatic approximation (to be described below), many  accurate results 
are obtained, including some such as current, and open-system phase boundaries, 
which appear to be exact in the large-system limit.

Section~\ref{sec:pbt} below gives the mean-field/adiabatic theory. 
In Section~\ref{sec:pbc} we investigate the TASEP with periodic boundary conditions;
in Sec.~\ref{sec:obc}, we examine open-boundary TASEP systems in 
the following phases: (a) maximal-current, (b) 
low-density, (c) high-density,  and (d) on the coexistence line. 
Finally, in Sec.~\ref{sec:conc}, concluding remarks are made.

\section{Preliminaries and Basic Theory}
\label{sec:pbt}

\subsection{Preliminaries}
\label{prelim}

We start by imposing periodic boundary conditions (PBC) for the TASEP at the ends 
of the chain, thus the total number of particles is fixed. For a uniform system 
in the steady state, the local average density at all sites coincides with the 
position-averaged particle density $\langle \rho\rangle$ (also to be denoted
below by $\rho$, wherever no chance of a misunderstanding arises)~.

Although this is a discrete model, we denote positions along
the lattice by a continuous variable $x$, thus (with the lattice parameter
being of unit length), the bond labelled by $x$ connects sites $x-\frac{1}{2}$ 
and $x+\frac{1}{2}$~. The use of a continuum description is consistent
with our emphasis throughout the paper on results applying in the infinite-system 
limit.

We consider a linearly-varying hopping rate; although
the theory developed in Subsection~\ref{mf} below applies to a general
position dependence (provided some rather general smoothness assumptions
are valid), this constant-gradient case will be our choice of concrete
application in the subsequent sections.  
For a system of size $L$, we take
\begin{equation}
p(x)=p_0 + \theta\,\frac{x}{L}\ ,\qquad -\frac{L}{2} \leq x \leq \frac{L}{2}\ ,
\label{eq:grad}
\end{equation}
where 
$\theta$ denotes the intensity of the hopping-rate gradient;
we keep $p_0=1/2$ henceforth.

\begin{figure}
{\centering \resizebox*{3.3in}{!}{\includegraphics*{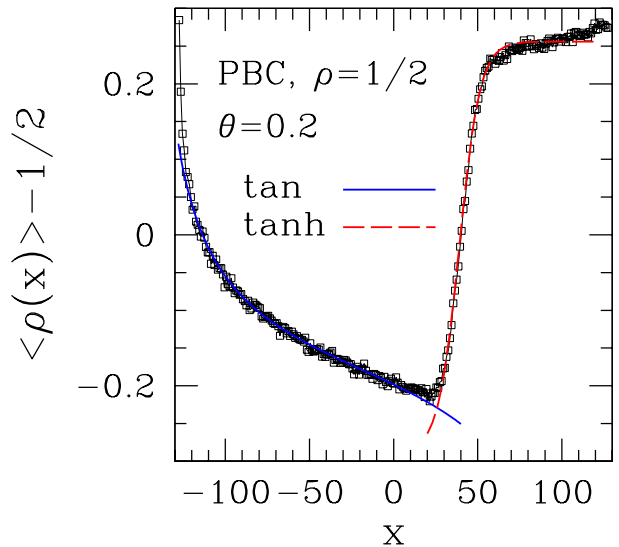}}}
\caption{(Color online) Points give steady-state density profile for TASEP with 
hopping-rate gradient, for  
periodic boundary  conditions, $\langle\rho \rangle=1/2$, 
lattice size $L=256$, and $\theta=0.2$ 
(see Eq.~(\protect{\ref{eq:grad}})). Full lines are fits to forms inspired
by the theory of randomly-disordered systems (see text). 
} 
\label{fig:prhoth02}
\end{figure}

The effect of the hopping-rate gradient, given by Eq.~(\ref{eq:grad}), on local
densities is rather remarkable, as illustrated in Figure~\ref{fig:prhoth02}~. 

A schematic interpretation of the profile shape displayed in 
Figure~\ref{fig:prhoth02}
can be provided as follows, using ideas from previous treatments 
of the quenched
random-bond version of the TASEP~\cite{tb98,hs04,dqrbs08}. For the TASEP
with uniform rates $p$, it is known~\cite{derr98,sch00,derr93,be07}
that, for currents greater or less than
$J_c (p)=\frac{p}{4}$ the steady state phases are characterized by density profiles
which are either: monotonically decreasing, $\langle \rho (x)\rangle -\frac{1}{2} =
-q\,\tan q(x-x_0)$ (high-current phase) or monotonically increasing,
$\langle \rho (x) \rangle -\frac{1}{2} =k\,\tanh k(x-x_0)$ (kink-like, 
low-current phase).
Here, $q$ and $k$ are characteristic inverse lengths such that
$q^2=-k^2=(J-J_c)/4p$~\cite{derr93,hs04,mukamel}, where $J$ is
the steady-state current; the profile forms result from the fact that $J$
is constant throughout the system. This latter fact has strong bearing on 
the local shape of density profiles in the quenched 
random-bond case:
in regions with
weak (strong) bonds, i.e. bonds with low (high) hopping probability $p_w$ ($p_s$),
$J$ can be larger (smaller) than the local critical current $J_c(p_w)$
($J_c(p_s)$), in which case the profile is of high-current (low-current) type.
With $\theta >0$ in Eq.~(\ref{eq:grad}), the features shown
in Figure~\ref{fig:prhoth02} appear roughly consistent with the theoretical
framework just sketched. However, we shall see from the full treatment
developed in Subsection~\ref{mf} below that, although the concepts of high- and 
low-current phases still persist here, their effects are strongly modified by 
factors specific to the present case.
In particular, the separation in space of the two phases is actually very
close to the left boundary in Figure~\ref{fig:prhoth02}, not where the tan and
tanh functions join in the fit shown in 
that same Figure.
This is because 
the actual profiles involve tan and tanh functions
with spatially varying "envelope" factors (see Subsection~\ref{mf}).
Many new features will be seen to arise from the "registration" in space of the
envelope, i.e. its position in the system; as we shall see, the location of the 
envelope relative to the region of weakest bonds is set by the current.

From the conjunction of PBC with the form of $p(x)$ 
given in Eq.~(\ref{eq:grad}), one sees that particles find a hopping-rate 
discontinuity of amplitude $-\theta$ as they  jump across the chain's endpoint.
Although, from  elementary considerations, PBC impose continuity of $\rho$ 
across the gap, it is important to emphasize that the kink-like profile seen in
Figure~\ref{fig:prhoth02} is not an artifact brought about by the discontinuity
just mentioned. As we shall see in the following, kinks may (or may not)
be present with PBC. Their existence, or lack thereof, depends on combinations 
of $\rho$ and $\theta$ according to mechanisms described by our theory. 

One should note that, if the sign of $\theta$ is reversed in Eq.~(\ref{eq:grad}), 
the plot of $\langle\rho \rangle-1/2$ versus $x$  simply gets 
point-reflected relative to the origin.

The steady-state currents in the type of system studied here also differ
markedly from their uniform counterparts. We recall that, for the latter with PBC,
the relationship between current $J_0 \equiv J(\theta=0,p,\rho)$ and density is
\begin{equation}
J_0=p\,\rho\,(1-\rho)\ ,
\label{eq:j0}
\end{equation}
where $p$ is the uniform hopping rate. 
Eq.~(\ref{eq:j0}) is one example of relationships and quantities which mean field 
factorization gives exactly~\cite{derr93,mukamel}, and whose generalization
for non-uniform rates is also exactly given by the generalized mean field 
theory developed here, as we shall see.
  
\begin{figure}
{\centering \resizebox*{3.3in}{!}{\includegraphics*{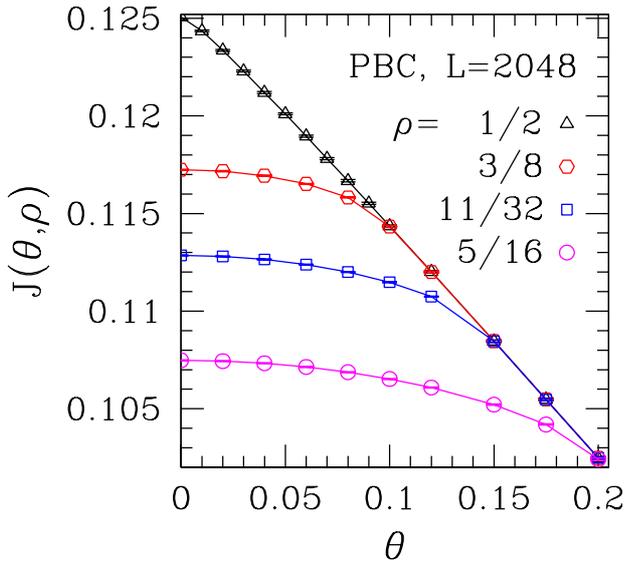}}}
\caption{(Color online) Steady-state currents $J$ against gradient intensity $\theta$
for a system with $L=2048$ and PBC, for densities as shown. Each point is an average
taken over $100$ independent samples, each in its turn containing
$1.2 \times 10^6$ successive steady-state configurations. Error bars are smaller than
symbol sizes.
} 
\label{fig:jth2048}
\end{figure}

For now, we restrict ourselves to $\rho \leq \frac{1}{2}$~. The question
of whether or not the $J-\rho$ diagram here displays the same symmetry, relative to
$\rho=\frac{1}{2}$, as  that for the uniform case
will be discussed later, with help of the theory developed in Subsection~\ref{mf}~.    
The effects on the current of a position-dependent $p(x)$ given by Eq.~(\ref{eq:grad})
are shown in Figure~\ref{fig:jth2048}, for various densities, all of them not
far removed from $\rho=1/2$. It is seen that as $\theta$ increases, the 
$J-\theta$ relationship becomes independent of $\rho$ for an increasingly
broad range of densities, following the same nearly linear  
form as that of a system with
$\rho=1/2$. In other words, for fixed $\theta$ a plateau develops around $\rho=1/2$
in the $J-\rho$  diagram, whose width increases with $\theta$~. 
Again, a similar effect is seen in TASEP with quenched randomness~\cite{tb98,hs04}~.  

\subsection{Mean-field theory}
\label{mf}

For uniform $p$, the Burgers equation~\cite{fns77,vBks85,ks91,kk08}, 
linearized via the Cole-Hopf transformation~\cite{hopf50,cole51} 
gives the general time-dependent mean field solution, analogous to
a superposition of moving solitons, corresponding to
waves in the linearized system, with
possibly complex wave-vectors. Real and imaginary wave
vectors distinguish the two general soliton-like steady states which,
because of particle conservation, are uniform-current ones.  These steady
states correspond to phases of maximal current ($J \simeq J_c=p/4$) or low current
($0 \leq J<J_c$, and low or high density); the square of the wave vector is proportional
to $(J/J_c -1)$. For the special case of PBC, 
the two steady states become states of uniform density, while for open boundary 
conditions the steady state profiles are of tan and tanh form.

For space-dependent $p(x)$, the solution given below (for general time-dependence
and then steady state) uses an adiabatic generalisation of constant-$p$ ideas.

We start from the continuity equation:
\begin{equation}
\frac{\partial \rho}{\partial t}=-\frac{\partial}{\partial x}\,J(x)\ ,
\label{eq:cont}
\end{equation}
with (using a mean field factorization)
\begin{equation}
J(x)=p(x)\,\rho (x-\frac{1}{2})\,\left(1-\rho (x+\frac{1}{2})\right)\ ;
\label{eq:jdef}
\end{equation}
defining $\sigma$ via
$\rho \equiv\frac{1}{2}(1+\sigma)$, Eq.~(\ref{eq:cont}) becomes, upon
taking the continuum limit on Eq.~(\ref{eq:jdef}):
\begin{equation}
2\,\frac{\partial \sigma}{\partial t}=\frac{\partial}{\partial x}\,\big\{
p(x)\,[\sigma^2(x)+\frac{\partial\sigma}{\partial x}-1\,]\big\}\ .
\label{eq:contsig}
\end{equation}
Using the Cole-Hopf transformation~\cite{hopf50,cole51}, we introduce the
auxiliary variable $u$ via $\sigma=\partial \ln u /\partial x$, in terms of
which, after a partial integration with respect to $x$,
Eq.~(\ref{eq:contsig}) turns into the linear form:
\begin{equation}
2\,\frac{\partial u}{\partial t} -f(t)\,u=p(x)\,\left\{\frac{\partial^2 u}
{\partial x^2}-u\right\}\ ,
\label{eq:c-h}
\end{equation}
where $f(t)$ is the integration "constant". Writing $u=X(x)\,T(t)$, one has
\begin{equation}
\frac{2}{T} \frac{dT}{dt}-f(t)=p(x)\left\{\frac{1}{X} \frac{d^2X}{dx^2}-1\right\}
\equiv -\omega\ ,
\label{eq:c-h2}
\end{equation} 
whence
\begin{equation}
T=\exp (-\frac{1}{2}\omega t +F(t))\quad{\rm with}\quad 
\frac{dF}{dt}=\frac{1}{2} f(t)\ .
\label{eq:sig_ell}
\end{equation}
Putting 
\begin{equation}
(\mu_\omega(x))^2 \equiv 1-\frac{\omega}{p(x)} \ ,
\label{eq:mu}
\end{equation}
and  making the {\em ansatz} $X=e^{\gamma (x)}$, one gets
\begin{equation}
\frac{d\gamma}{dx}= \pm \mu_\omega(x)\ ,
\label{eq:adiab}
\end{equation}
provided $d\mu_\omega(x)/dx \ll (\mu_\omega (x))^2$ ({\em adiabatic approximation}).
In this limit $X=e^{\pm\gamma_\omega(x)}$, with $\gamma_\omega(x) =
\int^x \mu_\omega (x)\,dx$. Thus,
\begin{equation}
XT = \exp (\pm\gamma_\omega(x)-\frac{1}{2}\omega t +F(t))\ .
\label{eq:XT_ad}
\end{equation}
The general solution  for $u(x,t)$ is
\begin{equation}
u=\sum_\omega A_\omega \cosh \left( \gamma_\omega (x) -\gamma_\omega (a_\omega)
\right)\,e^{-\frac{1}{2} \omega t +F(t)}\ ,
\label{eq:u_ad}
\end{equation}
where the $A_\omega$ and $a_\omega$ are arbitrary  constants. Finally, in
terms of $\sigma(x,t)$:
\begin{equation}
\sigma(x,t)=\frac{\sum_\omega A_\omega\,\mu_\omega (x)\sinh \left( \gamma_\omega (x) 
-\gamma_\omega (a_\omega)\right)\,e^{-\frac{1}{2} \omega t}}
{\sum_\omega A_\omega \cosh \left( \gamma_\omega (x) -\gamma_\omega (a_\omega)\right)
\,e^{-\frac{1}{2} \omega t}}\ ,
\label{eq:sig_ad}
\end{equation}
in the mean field/adiabatic approximation.

The following comments are in order:

\noindent 
(i)\ If we take a single component in Eq.~(\ref{eq:sig_ad}) the
$e^{-\frac{1}{2}\omega t}$ factor cancels and we are left with a steady state
solution:
\begin{equation}
\sigma (x)=\mu_\omega (x)\tanh\left(\gamma_\omega (x)-\gamma_\omega (a)\right)\ .
\label{eq:sing_comp}
\end{equation}
When the validity criterion for the adiabatic approximation applies, this state
is associated with the current
\begin{equation}
 J = \frac{1}{4}\,p(x) (1-\mu_\omega (x)^2)=\frac{1}{4}\omega\ ,
\label{eq:j_sing_comp}
\end{equation}
[$\,$using Eqs.~(\ref{eq:jdef}),~(\ref{eq:mu}), and~(\ref{eq:sing_comp})$\,$], 
which is constant as necessary for the steady state.

\noindent
(ii)\ In the $t$-dependent general form Eq.~(\ref{eq:sig_ad}), 
each sum evolves for long times into a
single component, which is the one having the least $\omega$, corresponding to the
steady state, i.e., $\omega=4J$ , by (i).

\noindent
(iii)\ At long but not infinite times the sums in  Eq.~(\ref{eq:sig_ad}) are 
dominated by the terms
with the smallest $\omega$'s. Then the denominator, whose logarithmic derivative
gives $\sigma$, becomes a combination of the steady state component and a wave
packet whose group velocity $v(x)$ can be obtained by a straightforward adiabatic
generalisation of standard procedures, using the analogue $d\mu_\omega (x)/dx$ of
the wave vector. The result is, generally,
\begin{equation}
v(x) =\pm p(x)\left(1-\frac{\omega}{p(x)}\right)^\frac{1}{2}\ ,
\label{eq:v(x)}
\end{equation}
becoming $v(x)= \pm p(x)^\frac{1}{2}(p(x)-4J)^\frac{1}{2}$ for the kink dynamics in
the late-time approach to the steady state.

In what follows we shall be mostly concerned with the steady state, so the
following distinctions and details may be helpful. In Eq.~(\ref{eq:j_sing_comp}),
\begin{equation}
J_c(x) \equiv \frac{1}{4}\,p(x)
\label{eq:jc(x)}
\end{equation}
acts like a local critical current, since the sign of $J-J_c$ determines whether
$\mu(x)$ there is real or imaginary and, consequently, whether the profile 
in Eq.~(\ref{eq:sing_comp}) involves a tanh or tan function. 
This is a generalization of the case  with space-independent rate $p$, 
where $J_c=p/4$ is the maximal current, associated to flat or $\tan$
profiles, while low currents $J < J_c$ exhibit $\tanh$ profiles. 

For the space-dependent $p(x)$
the most important new features are the $x$-dependence of $J_c(x)$,
the $J$-dependent location ($x_0$) of the division between phases, and the
occurrence of the space-dependent amplitude function $\mu(x)$ in the 
profile, Eq.~(\ref{eq:sing_comp}). Where it is necessary, 
to avoid confusion, we distinguish the possibilities by
using, in place of $\mu(x)$, the specific real functions $k(x)$, $q(x)$ defined by
\begin{eqnarray}
k(x)=\mu(x)=(1-4J/p(x))^\frac{1}{2}=(1-J/J_c(x))^\frac{1}{2}\ , 
\nonumber \\ J<J_c(x)\ ;\quad
\nonumber \\
q(x)=i\,\mu(x)=(4J/p(x)-1)^\frac{1}{2}=(J/J_c(x)-1)^\frac{1}{2}\ ,
\nonumber \\  J>J_c(x)\ .\quad 
\label{eq:kx_qx}
\end{eqnarray}
Then
\begin{eqnarray}
\sigma(x)=k(x)\,\tanh (K(x)-K(a)),\quad     J<J_c(x)\ ;\
\nonumber \\
\sigma(x)=-q(x)\,\tan (Q(x)-Q(b)),\quad     J>J_c(x)\ ,\
\label{eq:kx_qx2}
\end{eqnarray}
where $K(x)=\int^x k(x)\,dx$ and $Q(x)=\int^x q(x)\,dx$.
For $x$-dependent rates, the $\tan$ form can only apply in at most a very limited
region (of size set by the weakest rates). This is because the $\tan$ function in
$\sigma(x)$ diverges, violating the physical requirement on the local density,
$|\sigma(x)| \leq 1$, unless its argument $Q(x)$ is limited to a range less than
$\pi$.  So the $\tanh$ form will actually account for most of the profile. 
If the integration
constant $a$ is inside the system the change of sign of the argument of the $\tanh$
function at $x=a$ corresponds to a kink there. For the $\tanh$, $k(x)$ acts like an
envelope, and its crucial effects in distinguishing scenarios and phases partly
relate to its registration, for which the $\tan$ part of the profile 
can play a dominant role.

\section{Steady State with PBC}
\label{sec:pbc}

\subsection{Introduction}
\label{th}

For the non-uniform system with $x \in [-L/2,L/2]$, PBC
impose the constraint on  $\sigma(x)$:
\begin{equation}
\sigma \left(-L/2\right) = \sigma \left(L/2\right) \ .
\label{eq:pbcdef}
\end{equation}
In addition to this, in order to fix arbitrary constants and determine the
steady state current $J$ and profile $\sigma (x)$, we need also to specify the
average density $\langle\rho\rangle$, in the equation
\begin{equation}
2\,(\langle\rho \rangle-\frac{1}{2}\,) = \langle\sigma \rangle =
\frac{1}{L}\,\int_{-L/2}^{L/2} \sigma(x)\,dx\ . 
\label{eq:rho_av}
\end{equation}
With the mean field/adiabatic approximation this becomes 
\begin{eqnarray}
\langle\sigma \rangle =
\frac{1}{L}\,\int_{-L/2}^{L/2} \mu (x) \tanh\left(\gamma (x)-\gamma (a)\right)\,
dx = \nonumber\\
=\left[\ln\cosh\left(\gamma (x)-\gamma (a)\right)\,\right]_{-L/2}^{L/2}\ .
\label{eq:rho_av2}
\end{eqnarray}
Here we used $\mu=d\gamma/dx$, and have reverted to non-specific notation,
not distinguishing $\tanh$ or $\tan$ (nor $\cosh$ or $\cos$). We will later 
have to verify that the criterion for use of the adiabatic approximation is satisfied.

From the general formulation above, $\mu$ and hence $\gamma$ are related to the
current $J$; it and the other parameter $a$ (the kink position in the case of real
$\gamma(a)$) are determined in terms of $\langle\rho \rangle$ by
Eqs.~(\ref{eq:rho_av}) and~(\ref{eq:rho_av2}) (for large
systems, the kink position will be sharp when the adiabatic approximation is 
satisfied)~.

\subsection{Rate gradient}
\label{rg}

From now on we deal with the specific case of 
linearly-varying $p(x)$ given in Eq.~(\ref{eq:grad})~.
With PBC and $\theta \geq 0$, one gets in the adiabatic
approximation, with the help of Eqs.~(\ref{eq:mu}), (\ref{eq:adiab}), 
and~(\ref{eq:j_sing_comp}):
\begin{equation}
\mu(x)=\left[1-\frac{4J}{p(x)}\,\right]^{\frac{1}{2}}=\left[\frac{X}{X+c}\,
\right]^{\frac{1}{2}}\ ,
\label{eq:lin_mu}
\end{equation}
and
\begin{eqnarray}
\gamma=\int^x \mu (x)\,dx=[X(X+c)]^{\frac{1}{2}}- 
c \tanh^{-1} \left[\frac{X}{X+c}\right]^{\frac{1}{2}}
\nonumber \\ \equiv {\widetilde K}(X)\ ,\qquad\quad
\label{eq:lin_gamma}
\end{eqnarray}
where
\begin{eqnarray}
X=x-x_0\ ,\qquad\qquad\qquad\qquad\quad \nonumber \\
x_0=(8J-1)\,\frac{L}{2\theta} \equiv -\lambda\,\frac{L}{2}\ ;\qquad\qquad \nonumber \\
c=\frac{4JL}{\theta} =x_0 + \frac{L}{2\theta}=\frac{L}{2}\,\left(\frac{1}{\theta}
-\lambda\right)\ .
\label{eq:x0def}
\end{eqnarray}
$x_0$ corresponds to the place where $\mu(x)$ vanishes, hence to the position of the
apex of the envelope function $\pm |\mu(x)|$, i.e., where $\mu (x)$ 
[$\,$and $\gamma(x)\,$] 
cross over between real and imaginary values $k(x)$ or $-i\,q(x)$
[$\,$and $K(x)$ or $-i\,Q(x)\,$]. Subsequently explicit forms will be needed,
particulary for $\gamma$  for the real case, and it will then be convenient
to use both $K$ and (real) $\widetilde K$, where 
\begin{equation}
K(x)={\widetilde K}(X)\ ,
\label{eq:KxKX}
\end{equation} 
with $X=X(x)=x-x_0$, and where $\widetilde K$ is as in Eq.~(\ref{eq:lin_gamma}).
$x_0$ also corresponds to the place where $J$
is equal to the local critical current; this plays a central role in the
discussion. For graphical illustrations, refer to Figure~\ref{fig:varprof02}
in subsection~\ref{nr} below.
$c$ is a characteristic length related to the rate gradient.
$\lambda$, the ratio of $x_0$ to $-L/2$, conveniently distinguishes 
scenarios, and parametrizes analytic expressions, particulary in  
the $L \to \infty$ limit.

\subsection{Scenarios for steady state behavior}
\label{sc}

We next discuss the character and location of steady state phases, 
and relationships to positions of the "envelope" and kinks.
The generalised maximal current and low current phases of the system turn out to
be described by two scenarios, I and II, as follows.

For the rate gradient case with $\theta>0$ ($\theta<0$ has dual character),
the smallest $p(x)$ is at the left-hand side edge, giving a severe bottleneck there. 
As we shall see in the following, this has the consequence that the current $J$ 
adjusts itself 
in such a way that the apex position $x_0 = -\lambda L/2$ turns out to be
either: (I) near the left boundary, but still inside the system, or (II) to the left
of the left boundary. These give, respectively:

\noindent
Scenario I:\ $\lambda \lesssim 1$. Here the $\tan$ function applies near the 
left edge and its spatial extent $\Delta x$ is limited by the condition 
$\Delta Q(x) =q(x)\,\Delta x < \pi$.
Since $q(x)$ is related to the difference $J - J_c(x)$ of the steady state current 
$J$ from its local critical value, this condition also limits $J$ as well as the
position, $x=x_0$, where $J - J_c(x)$ vanishes.

\noindent
Scenario II:\ $\lambda > 1$. Here only the $\tanh$ function applies inside the system.

The two scenarios become very evident in the "family" of profiles corresponding
to all possible average densities $\langle \rho \rangle$, for 
PBC and a given $\theta$
(see the numerical results in Figure~\ref{fig:varprof02}). 

Scenario I corresponds to a common envelope (nearly parabolic in shape, see
Eq.~(\ref{eq:lin_mu})) and
applies for an intermediate range of $\langle \rho \rangle$'s (not very far from  
$1/2$). It is consistent with a fixed position $x_0$ of the apex of the envelope, 
close to the left hand boundary. It is (through Eq.~(\ref{eq:x0def})) 
consistent with an observed constant ($\langle\rho \rangle$-independent) 
plateau current $J$, about $(1/8)(1-\theta)$. Near the left
boundary there is a small region of $\tan$  profile, and everywhere else the
profile approaches the $\tanh$ form (including the kink).

Scenario II, applying for larger $| \langle\rho \rangle-1/2|$, has profiles 
not near a common envelope, corresponding to varying apex position; 
indeed, in this case $x_0$ is outside of the system (to the left of the 
left boundary) and the profile is entirely of $\tanh$  type. In this scenario 
the currents depend on $\langle \rho \rangle$.

These scenarios and related phenomena can be quantitatively explained using the
mean field adiabatic formulation, except near the envelope apex if that lies
inside the system. This is because the apex is where $\mu(x)$ vanishes,
i.e., where the adiabatic approximation fails utterly (see the validity criterion,
below Eq.~(\ref{eq:adiab})). 
To the right of the apex, where $J<J_c(x)$, the adiabatic approximation
is valid for $X> c^{\,1/3}$, so the adiabatic form 
$\sigma_R=k(x)\,\tanh(K(x)-K(a))$ applies; similarly, in the region
to the left of the apex, $J>J_c(x)$, and the adiabatic form
$\sigma_L=-q(x)\,\tan(Q(x)-Q(b))$ is valid for $X<-c^{\,1/3}$~.
Between these a (nonadiabatic) form $\sigma_C \propto (x+{\rm const.})^{-1}$
is adequate. So the profile can be a piecewise combination of
$\sigma_L$, $\sigma_C$, and $\sigma_R$, except for scenario II, where only $\sigma_R$
applies.

In all cases, for the integral in Eq.~(\ref{eq:rho_av}) for ($2\langle\rho \rangle-1$) 
it turns out that at large $L$
the contribution from $\sigma_R$ dominates, and it alone gives the $L \to \infty$
value. This is because of the limitation of the range of the $\tan$ function in
$\sigma_L$, to prevent its divergence, and of the range [$\,\sim~c^{1/3}~\propto~
L^{1/3}\,$] of $\sigma_C$. This makes their contributions
 to the integral less than that from $\sigma_R$ by a factor which vanishes as $L$
increases.

Note that, quite generally, the limitation of $\sigma_L$ requires $x_0$ to satisfy
$x_0-(-L/2) <\pi/q(-L/2)$; in the limit $L \to \infty$ this restricts the 
variable $\lambda$ defined above to the two possibilities $\lambda =1$
(envelope apex very near the left boundary) or $\lambda >1$ (apex [well] outside). 
These are respectively Scenarios I and II, whose details are now exhibited.

\subsubsection{Scenario I}
\label{scI}

In this case, where $\lambda =1$, we investigate its quantitative character 
and which values of $\langle \rho \rangle$ and $J$ are consistent with it.
Firstly, from Eq.~(\ref{eq:x0def}), $\lambda=1$ makes $J=(1/8)(1-\theta)$.
For possible values of $\langle \rho \rangle$, we consider 
Eqs.~(\ref{eq:rho_av}) and~(\ref{eq:rho_av2}).
We chose the integration constant $a$ so that $x=a$ is the center of the kink.  
If the kink is inside the system, the further it is to the right the 
smaller will be the integral, and the associated $\langle \rho \rangle - 1/2$. 
There is clearly a least $\langle \rho \rangle$ in Scenario I, applying when
the kink is as far to the right as it can be (consistent with PBC). But
Scenario II allows displacement of the {\em envelope} to the left 
($\lambda > 1$), and with fixed kink position this
affects the value
of the integral, since the more the envelope is displaced to the left, the
larger will be the amplitude $k(x)$ of the $\tanh$ at any particular $x$ inside the
system.

So, small  $|\langle \rho \rangle - 1/2|$ can be achieved with envelope 
apex near the left hand boundary, by adjusting the kink position (Scenario I),
while $\langle \rho \rangle$ nearer $0$ or $1$
needs a large displacement (${\cal O}(L)$) of the envelope to the left, 
corresponding to $\lambda > 1$ (Scenario II).

For illustration, consider the special case  $\langle \rho \rangle = 1/2$ for which
the numerical profile is actually shown in Fig.\ref{fig:prhoth02}. 
In the Figure it is evident that the required zero value of the integral 
between the profile curve and the $x$-axis is achieved by having the abrupt 
rise of the curve, corresponding to the kink, where it is. To the right(left) 
of the kink the curve follows the upper (lower) branch of the envelope function 
($k(x)$ is monotonically increasing). At the extreme left is the region 
around $x_0$ (necessarily small) where the $\tanh$
has become $\tan$; its near divergence makes it easily able to match the PBC
requirement. Thus one sees, in retrospect, that the fit shown in
Fig.\ref{fig:prhoth02} is in fact quite misleading.
 
Scenario I is consistent as long as the kink stays within the right
boundary of the system. Then, $\sigma(L/2)$ at that boundary is positive and the 
PBC requiring $\sigma(-L/2)$ to have the same 
positive value can be readily satisfied, as the $\tan$ form needs only a 
very small adjustment of its argument (within $ \approx \pi$) to achieve this. 
At the same time the spatial range in which the
$\tan$ form applies has to be very small to prevent unphysical $\sigma_L$'s~.
Of course $\sigma_L$ and $\sigma_C$, and the relationship of their
constants to $a$ are needed to complete the determination of the profile.

This discussion is easily generalized and made more quantitative by using the
integration result in Eq.~(\ref{eq:rho_av2}), 
with the appropriate real version $K(x)$ of $\gamma(x)$,
together with the fact that when 
the kink at $X=a-x_0 \equiv A$ lies well 
inside the system ${\widetilde K} (X)- {\widetilde K} (A)$ is large 
(${\cal O}(L)$) at both limits, but of opposite signs. 
Further, the kink width ($\equiv w$), such that the argument of $\tanh$ 
in Eq.~(\ref{eq:rho_av2}) changes by 
${\cal O}(1)$ between $x=a \pm w/2$,
is $w \approx [d{\widetilde K}(A)/dA]^{-1}=[(A+c)/A]^{1/2}$, which is ${\cal O}(1)$
for $a={\cal O}(L)$, except near $x_0$ where $w$ diverges.
Hence the integration result is, in the limit of large $L$,
\begin{equation}
2\,(\langle\rho \rangle-\frac{1}{2}\,) = \frac{1}{L}\,
\left({\widetilde K} (L)-2{\widetilde K} (A)\right)\ ,
\label{eq:rho_av3}
\end{equation}
where 
\begin{equation}
{\widetilde K}(X)=[X(X+c)]^{1/2}- c\tanh^{-1} \left[\frac{X}{X+c}\right]^{1/2}\ .
\label{eq:rho_av4}
\end{equation}

\begin{figure}
{\centering \resizebox*{3.3in}{!}{\includegraphics*{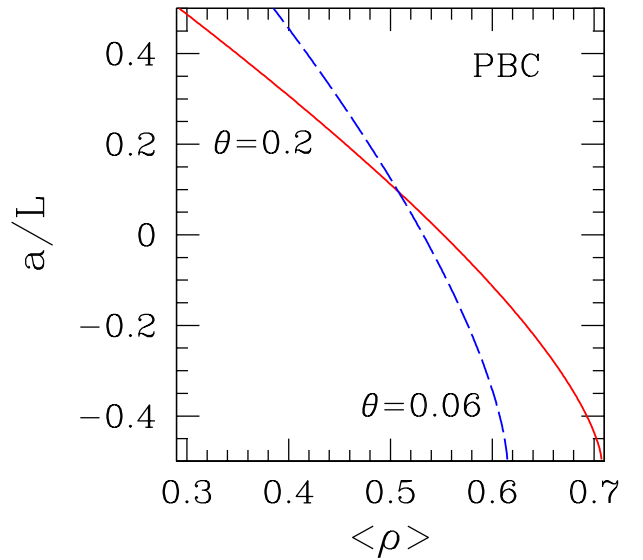}}}
\caption{(Color online) Kink position $a/L$ against density $\langle \rho 
\rangle$ [$\,$see Eqs.~(\protect{\ref{eq:rho_av3}}) 
and~(\protect{\ref{eq:rho_av4}})$\,$], for systems 
with PBC and rate-gradient values $\theta$ as shown. 
} 
\label{fig:kvsrho}
\end{figure}

For the special case $\langle\rho \rangle=1/2$ the kink position then has to be 
such that ${\widetilde K}(A)=(1/2){\widetilde K} (L)$ which, 
using the explicit form  of $\widetilde K$ [$\,$see 
Eqs.~(\ref{eq:KxKX}) and (\ref{eq:rho_av4})$\,$], gives 
$A/L=(1/2)+a/L \approx 0.61$ for $\theta=0.2$, consistent with the kink position in 
Figure~\ref{fig:prhoth02}. 
The general solution for the kink position against particle density 
is exhibited in Figure~\ref{fig:kvsrho}, for illustrative values of $\theta$.
Note that the range of values of $\langle \rho \rangle$ for which solutions are found
is symmetric relative to $\langle \rho \rangle =1/2,$ and gets broader with
increasing $\theta$ [$\,$see also Eq.~(\ref{eq:rhovsth}) below$\,$].

Larger values of $a$ are associated, through Eq.~(\ref{eq:rho_av2}), with 
$\langle\rho \rangle < 1/2$, up to the limit $a=L/2$ when the kink center 
is at the right boundary. Then the mean density takes the limiting value 
$\langle\rho \rangle_c$ such that
\begin{eqnarray}
(\langle \rho \rangle_c-\frac{1}{2}\,) = 
-\frac{1}{2L}\,{\widetilde K} (L)=\qquad\qquad\quad \nonumber \\
=\frac{1}{4} \{[2(1+\frac{1}{\theta})]^{1/2} -
(\frac{1}{\theta}-1)\tanh^{-1}\left[\frac{2\theta}{1+\theta}\right]^{1/2}\}
\ .\qquad 
\label{eq:rhovsth}
\end{eqnarray}
This marks the condition where the two scenarios meet, and will correspond to
the limit of a plateau region (in which, for $L \to \infty$, $J=\frac{1}{8}(1-\theta)$
applies) in the "fundamental" diagram relating $J$ with $\langle \rho \rangle$
and $\theta$.

At $\langle \rho \rangle > 1/2$ one finds equations identical in form to
Eq.~(\ref{eq:rho_av3}),~(\ref{eq:rho_av4}), and~(\ref{eq:rhovsth}), 
with $\frac{1}{2}-\langle \rho \rangle$ replacing $\langle \rho \rangle-\frac{1}{2}$. 
For small $\theta$, Eq.~(\ref{eq:rhovsth}) gives 
$\langle \rho \rangle_c -1/2 =-(\sqrt{2}/3)\,\theta^{\,1/2}$. 
Thus, the extent of the plateau in the $J-\langle \rho \rangle$ diagram 
vanishes as $\theta \to 0$. In this limit, for $\langle \rho \rangle$  
still within the plateau, one can show that  the height of the kink vanishes as
$\theta^{1/2}$. 

\subsubsection{Scenario II}
\label{ScII}

Scenario II applies at $\langle \rho \rangle$'s so small 
(for a given $\theta$) that the kink center is beyond the right boundary of the 
system (see, e.g., the curves for $\langle \rho \rangle=0.25$, $0.125$ in 
Figure~\ref{fig:varprof02}). Then the apex position $x_0$ of the $\tanh$
envelopes has to go outside of the system on the left, and there has to be a
small upturn in $\sigma$ at the extreme right of the system to satisfy the 
PBC, so
the start of the kink is just visible there in Figure~\ref{fig:varprof02},
and the kink center is actually beyond the right boundary. 
This means that the $\tanh$ profile applies throughout the system:
\begin{equation}
\sigma(x)=\sigma_R=k(x)\,\tanh(K(x)-K(a))\ ,
\label{eq:rhovsxR}
\end{equation}
where $K(x)$ is again as in  Eqs.~(\ref{eq:lin_gamma}) and~(\ref{eq:KxKX}).
We will now have $x_0<-L/2$ and $L/2 < a <L/2+w$, where $w$ is the kink 
width (of order $1$).

As discussed above, for  given $\theta$ specifying $\langle \rho \rangle < 
\langle\rho \rangle_c < 1/2$ [$\,$see Eq.~(\ref{eq:rhovsth})$\,$]
will lead to $x_0/L < -1/2$, so making $J <(1/8)(1-\theta)$ and $\lambda >1$.
As before, we use Eqs.~(\ref{eq:rho_av2}) and~(\ref{eq:lin_gamma}). 
But now, since $x_0<-L/2$, and for all 
$x$ in the system $x<a$, ${\widetilde K} (x)-{\widetilde K} (a)$ 
is at both limits negative (and large). So we have
[$\,$ignoring contributions to $2(\langle\rho\rangle-1/2)\,L$ of lower order in $L$
(from corrections to the adiabatic approximation, and from width of the kink), 
and the comparable small distance the center lies beyond the right boundary$\,$]:
\begin{equation}
2(\langle\rho \rangle-\frac{1}{2}) \approx 
\frac{1}{L} \{\,{\widetilde K} (\frac{L}{2}(\lambda+1))-
{\widetilde K} (\frac{L}{2}(\lambda-1))\,\}
\label{eq:rhovsKR2}
\end{equation}
where ${\widetilde K}(X)$ is as in  Eq.~(\ref{eq:rho_av4}).
This gives $\lambda$ in terms of $\langle \rho \rangle$ and $\theta$, 
for $\langle\rho \rangle$ less than the critical value, and hence provides the 
following current-density relation outside of the plateau region 
\begin{equation}
J=\frac{1}{8}\,(1-\lambda \theta)\ ,
\label{eq:jvsth2}
\end{equation}
with
\begin{eqnarray}
4\,(\frac{1}{2} - \langle\rho \rangle)= [(\lambda +1)(\frac{1}{\theta}+1)]^{\frac{1}{2}} 
-[(\lambda-1)(\frac{1}{\theta}-1)]^{\frac{1}{2}} -
\nonumber \\
-(\frac{1}{\theta}-\lambda)\,\{\tanh^{-1} \left[\frac{\lambda+1}{\theta^{-1}+1}
\right]^{\frac{1}{2}}
- \tanh^{-1}\left[\frac{\lambda -1}{\theta^{-1}-1}\right]^{\frac{1}{2}}\}\ .
\nonumber \\
\label{eq:rhovsth2}
\end{eqnarray}
A similar procedure applies for the complementary subcase, $(1-\langle \rho \rangle)
< \langle\rho \rangle_c < 1/2$, by particle-hole duality.

Eqs.~(\ref{eq:jvsth2}) and~(\ref{eq:rhovsth2}) can be combined to give $J$ as 
a function of $\langle \rho \rangle$ in Scenario II, for fixed $\theta$. 
The range of values of $\langle \rho \rangle$ for which physically acceptable
solutions are found is complementary to that limited by Eq.~(\ref{eq:rhovsth}), 
which marks the extremes of validity of Scenario I.

\subsubsection{Weak-bond interpretation of plateau current}
\label{w-b}

Before moving to numerical results, we introduce an additional
piece of mean-field theory which will be useful later.

As remarked above, a plateau current, similar to that predicted in Scenario I,
is found in the TASEP with random rates $p(x)$~\cite{tb98,hs04,dqrbs08}. 
There, an interpretation is given in terms of the current limitation
provided by the weakest bonds, $p_w$, which suggests that the "maximal" current
satisfies $J_{\rm max} \leq p_w/4$. The following generalization provides a direct
interpretation and confirmation of the result $J_{\rm max}=\frac{1}{8}(1-\theta)$
predicted for the plateau phase in Scenario I.

In the continuum mean field formulation, Eqs.~(\ref{eq:cont}), (\ref{eq:jdef}),
and~(\ref{eq:contsig}) give for all $x$:
\begin{equation}
J=\frac{1}{4}\,p(x)\,\{1 -\sigma^2(x) -\frac{\partial \sigma}{\partial x}\}\ .
\label{eq:jsig}
\end{equation}
The most limiting rate, occurring at $x=-L/2$, is $p_w=\frac{1}{2}\,(1-\theta)$,
so $J_{\rm max}$ is obtained from applying Eq.~(\ref{eq:jsig}) there. In Scenario I, 
with $\lambda=1$, the $\tan$ solution  Eq.~(\ref{eq:kx_qx2}) applies in that
region [$\,$i.e. $X = {\cal O}(1)\,$], which yields for the right-hand side of
Eq.~(\ref{eq:jsig}), using Eqs.~(\ref{eq:kx_qx}), (\ref{eq:lin_mu}), 
and ~(\ref{eq:x0def}):
\begin{equation}
\frac{1}{4}\,p_w\,(1+q^2)=\frac{1}{4}\,p_w\,\{1-\frac{X}{X+c}\}=
\frac{1}{4}\,p_w\,\{1+{\cal O}(\frac{1}{L})\}\ ,
\label{eq:jsig2}
\end{equation}
hence confirming the infinite-system maximal current $\frac{1}{8}\,(1-\theta)$.

\subsection{Numerical results}
\label{nr}

We considered lattices with $L=2^m$ sites, $8 \leq m \leq 13$.
A time step is defined as a set of $L$ sequential update attempts,
each of these according to the following rules: (1) select a site at random;
(2) if the chosen site, here denoted by $x$, is occupied and its neighbor to the 
right is empty, then (3) move the particle with probability $p(x)$. Thus, in the
course of one time step, some sites may be selected more than once for examination,
and some may not be examined at all. 

We have found that the time needed to attain steady-state flow varies roughly
with $L^{3/2}$, similarly to the uniform-rate case~\cite{dqrbs08}, for which
this is well-known~\cite{fns77,gwa,dhar}, and is in agreement with the
correspondence between the (uniform) 
TASEP and  evolution of a KPZ interface~\cite{ks91,kk08,kpz,meakin86}.

For density profiles, local densities were usually averaged over
snapshots (taken at appropriately long times) of $10^4$ independent samples.
For example, for $L=256$ we  found that steady state has been reached 
by time $t=10^4$ in most cases, except for points on the coexistence line
for open BCs (see Section~\ref{sec:obc}) where the approach to stationarity is
markedly slower. Although finite-size effects can be observed, they are
generally small and act towards making any kinks sharper, relative to system
size, without any qualitative change. Thus we can be confident that no 
significant physical features are missed by generally exhibiting
profiles corresponding only to $L=256$, as done here.

\begin{figure}
{\centering \resizebox*{3.3in}{!}{\includegraphics*{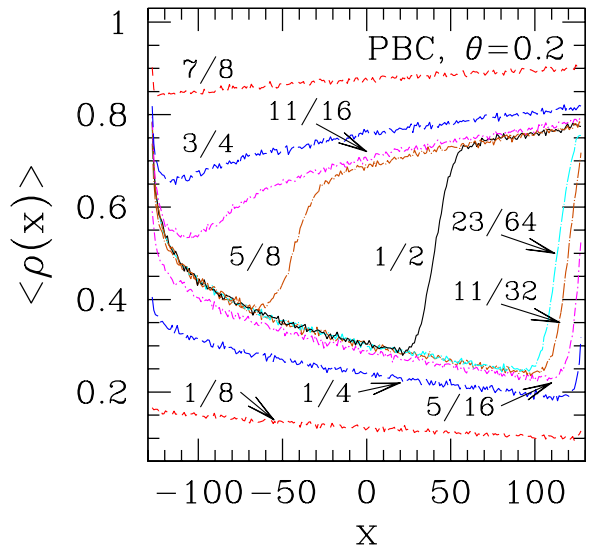}}}
\caption{(Color online) Steady-state local density profiles for system with $L=256$ and
PBC, $\theta=0.2$. Curve labels denote average particle densities.
} 
\label{fig:varprof02}
\end{figure}

Figure~\ref{fig:varprof02} shows steady-state density profiles for $\theta=0.2$,
which although still in the scaling regime is a relatively steep gradient.
The behavior is in full agreement with the theory developed in 
Subsection~\ref{mf}: (i) according to Scenario I, 
there is a common envelope, pinned to the left-hand extreme of the system, 
for intermediate densities roughly between
$0.3$ and $0.7$; (ii) within this range of densities a kink is present, whose 
location varies against $\langle \rho \rangle$ as predicted by 
Eqs.~(\ref{eq:rho_av3}) and~(\ref{eq:rho_av4}); (iii) for densities further removed  
from $1/2$, Scenario II takes over, and profiles follow either the lower 
branch of the envelope (with its $\langle \rho \rangle$- dependent displacement) 
with an incipient kink at the right boundary
(for $\langle \rho \rangle <1/2$), or the upper branch, in this case with
a narrow downward turn at the left edge in order to satisfy PBC 
($\langle \rho \rangle > 1/2$).

Envelope functions are already familiar in the profiles of
constant-rate asymmetric exclusion processes (e.g., on the coexistence
line), but they only involve new scenarios when their delimitation
of density profiles is space-dependent (as above or, e.g., in
asymmetric exclusion problems with Langmuir dynamics~\cite{pff04}). 

There are slight numerical discrepancies between predictions of Subsection~\ref{mf} 
and the data displayed in Fig.~\ref{fig:varprof02}, which exemplify the finite-size 
effects referred to above. For instance, according to 
Eqs.~(\ref{eq:rho_av3}),~(\ref{eq:rho_av4}), and~(\ref{eq:rhovsth}) [$\,$see also 
Fig.~\ref{fig:kvsrho}$\,$], for $\theta=0.2$ Scenario I
should hold for $0.293 \dots \leq \langle \rho \rangle \leq 0.707 \dots$.
However, the profile for $\langle \rho \rangle=5/16=0.3125$ already shows
some deviation from the common envelope. Overall, we have found that 
the quantification of finite-system corrections, together with
accurate analysis and extrapolation to the $L \to \infty$ limit, can best be
accomplished when dealing with steady-state currents, as shown in the following.

\begin{figure}
{\centering \resizebox*{3.3in}{!}{\includegraphics*{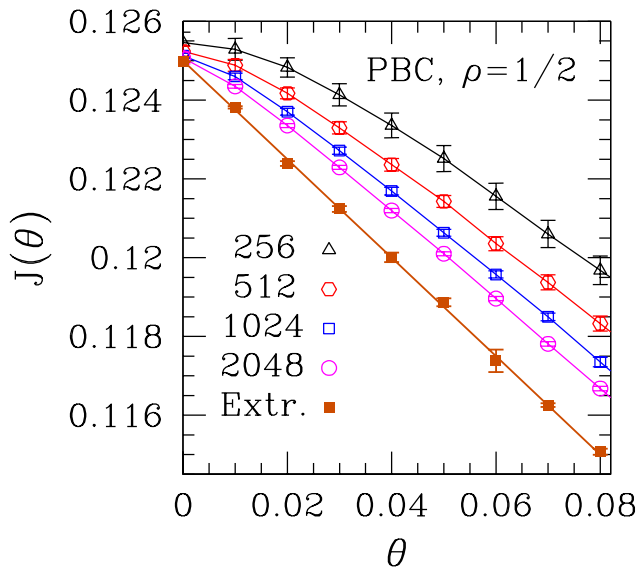}}}
\caption{(Color online) Steady-state currents $J$ against gradient intensity $\theta$
for $\rho=1/2$ and PBC, for system sizes as shown, plus extrapolated curve; 
for details of extrapolation, see text.
} 
\label{fig:psscurr}
\end{figure}

Evaluation of steady-state currents involved averaging over $N_s=100$ independent 
samples, for each of which $N_c= a_L\,L^{3/2}$ successive instantaneous 
current values were accumulated. We took $a_L \approx 130$ for $L=256$ and $512$,
and $\approx 13$ for larger $L$. The instantaneous current is $n_{\rm moves}/L$,
where $n_{\rm moves}$ is the number of particles which undergo successful move
attempts in the course of a unit time interval, i.e., $L$ stochastic site 
probings as defined above. As is well known~\cite{dqrbs96}, the width $\delta J$
of the distribution thus found is essentially independent of $N_s$ 
as long as $N_s$ is not too small, and varies as $N_c^{-1/2}$. With the parameters
as specified here, we managed to keep $\delta J$ well below the finite-size difference
between $J$  estimates for consecutive values of $L$ (for fixed $\theta$,
$\rho$). The relevance of finite-size effects for currents is illustrated
for $\rho=1/2$ in 
Fig.~\ref{fig:psscurr}, where $\theta$ is restricted to small values for clarity of
presentation; one can see that  the curvature present in finite-$L$ data is
essentially absent upon extrapolation to $L \to \infty$. 
We now discuss guidelines for extrapolation of  finite-system currents $J_L$
to their thermodynamic-limit value $J_\infty$. 

In line with general finite-size scaling ideas, 
we attempted single-power fits of our sequences of finite-$L$
current data with an adjustable finite-size scaling exponent $\psi$, 
for all available pairs $\theta$ and $\rho$. 
We denote by $\theta_c(\rho)$ the gradient intensity value above which
$J(\rho,\theta)$ becomes independent of $\rho$. 

So, $\theta <\theta_c(\rho)$
corresponds to Scenario II of Subsection~\ref{sc} above, while
$\theta >\theta_c(\rho)$ is associated with Scenario I.
Although $\theta_c$ still carries an
$L$-dependence (thus, e.g., the mergings of $J-\theta$ curves shown in 
Fig.~\ref{fig:jth2048} take place at slightly different locations for
$L \neq 2048$), it is a rather small effect compared to the overall range of
$\theta$-variation investigated.

Results were as follows:\par\noindent
(1)\ For $0 \leq \theta \lesssim \theta_c(\rho)$, $\psi \approx 1$\ (Scenario II);
\par\noindent
(2)\ For $\theta \gtrsim \theta_c(\rho)$, $\psi \approx 1/2$\ (Scenario I).
\par\noindent
In the immediate vicinity of $\theta_c(\rho)$, on both sides, we had
rather serious convergence issues, so there we generally resorted to fixing
$\psi=1/2$, for which the corresponding extrapolations fell in smoothly with
the remaining ones outside that interval. For region (2), we estimate
the uncertainty for $\psi$ to be of order $10\%$ at most.

Thus, for the extrapolated points
in  Fig.~\ref{fig:psscurr}, $\psi \in (0.45,0.55)$ was
found in all cases  except that corresponding to $\theta=0$, for 
which $\psi \approx 1$. The case of $\langle \rho \rangle =1/2$ shown in that
Figure is somewhat
exceptional in that, as remarked at the end of Subsection~\ref{scI} above, 
there the extent of validity of Scenario II corresponds only to the limit 
$\theta \to 0$.

\begin{figure}
{\centering \resizebox*{3.3in}{!}{\includegraphics*{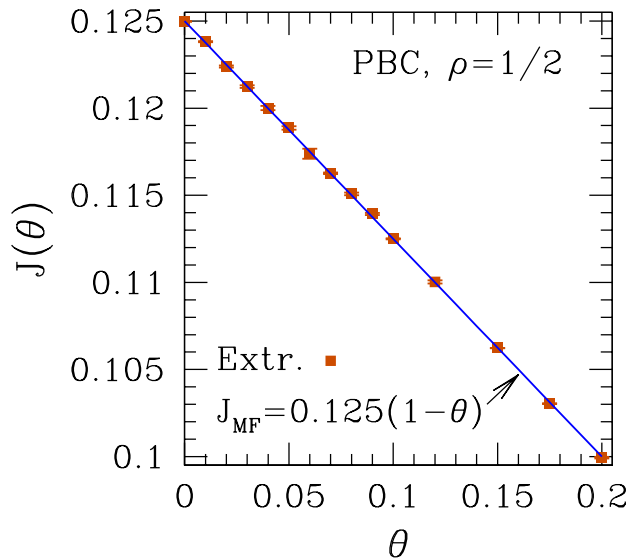}}}
\caption{(Color online) Points are extrapolated currents for system with
$\rho=1/2$, PBC. Full line is the mean-field approximation (see text).
} 
\label{fig:rho05ext}
\end{figure}

In Fig.~\ref{fig:rho05ext} below we present the set of extrapolated
currents for $\rho=1/2$, corresponding to $0 \leq \theta \leq 0.2$,
together with the mean-field prediction of a straight line 
$J_{\rm MF}(\theta)=\frac{1}{8}\,(1-\theta)$~ for Scenario I (see also the
weak-bond interpretation given in Subsec.~\ref{w-b}). 
The agreement is remarkable.

\begin{figure}
{\centering \resizebox*{3.3in}{!}{\includegraphics*{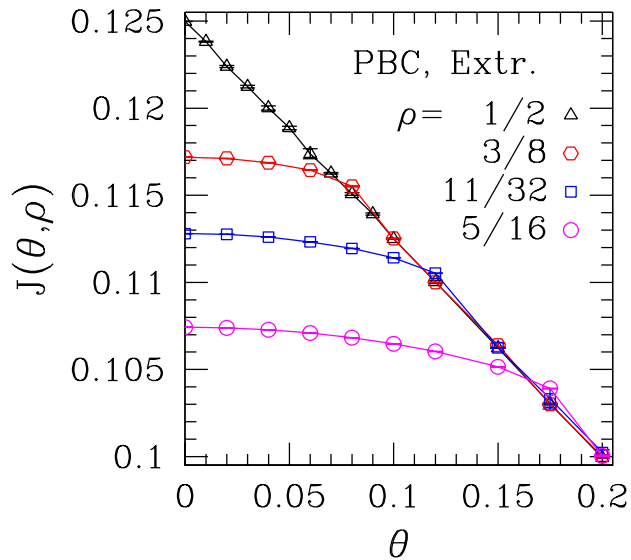}}}
\caption{(Color online) Extrapolated steady-state currents versus gradient
intensity for systems with PBC, and densities as shown. Note overshoots.
} 
\label{fig:jthext}
\end{figure}

Considering now the extrapolated currents for
$\langle \rho \rangle \neq 1/2$, one sees in Fig.~\ref{fig:jthext}
that the variation of $J$ against $\theta$ is generally much slower where
Scenario II holds. In the vicinity of $\theta_c(\rho)$, due to the convergence issues
mentioned above, we considered systems of sizes up to $L=8192$ (away from that region,
we found that using $L \leq 2048$ was generally enough to distinguish
a reliably smooth trend as $L^{-1} \to 0$). Upon extrapolation we found the small
overshoots shown in the Figure, which when translated to $J-\langle\rho \rangle$ 
diagrams for fixed $\theta$, would amount to reentrant behavior. 
For the largest deviation found,
corresponding to $\rho=5/16$ at $\theta=0.175$, one gets $J=0.1039(1)$,
while the value for $\rho=1/2$ at the same $\theta$ is $0.1030(1)$.
Although the average values differ by just under $1\%$, when converted in
terms of (estimated) uncertainties this difference is equivalent to nine error bars. 
So, this effect appears to be real.

The data in Fig.~\ref{fig:jthext} can be used to test Eq.~(\ref{eq:rhovsth}).
In order to do so, for fixed $\langle \rho \rangle <1/2$ one needs to establish 
the boundary between the ranges of validity of
Scenarios I and II, as given by numerical simulations. Due to the overshoots
just referred to, this task carries some ambiguity. For simplicity,
we assumed such location to be where the 
respective $J-\theta$ curve first crosses that for  $\langle \rho \rangle =1/2$,
upon increasing $\theta$. Fitting the data thus obtained to the form
$\frac{1}{2}-\langle \rho \rangle =a\,\theta^b$, one finds $a=0.471(4)$,
$b=0.51(1)$. These are to be compared, respectively, to
$a=\sqrt{2}/3=0.4714 \dots$, $b=1/2$, from the
small-$\theta$ expression of Eq.~(\ref{eq:rhovsth}) (see paragraph below 
that Equation).
Thus, the above assumption seems justified.

\begin{figure}
{\centering \resizebox*{3.3in}{!}{\includegraphics*{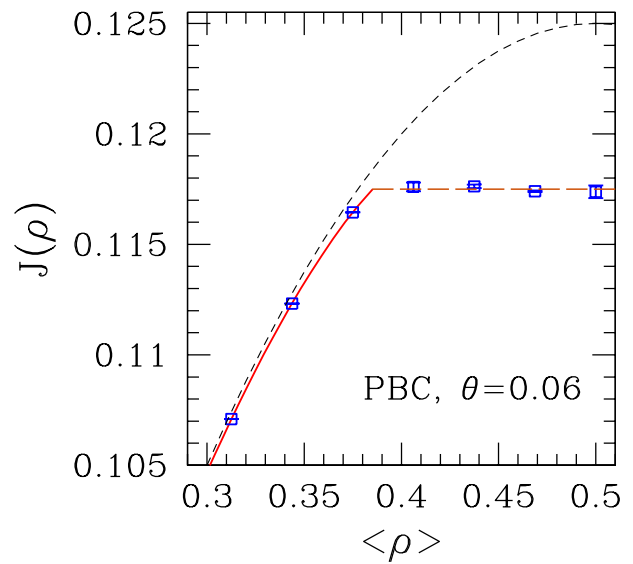}}}
\caption{(Color online) Points are extrapolated ($L \to \infty$) steady-state currents;
full line is $J-\langle \rho \rangle$ relationship from Eqs.~(\protect{\ref{eq:jvsth2}}) 
and~(\protect{\ref{eq:rhovsth2}}) [$\,$Scenario II$\,$]; long-dashed horizontal 
line is $J=0.1175$,
from Eq.(\protect{\ref{eq:x0def}}) with $\lambda=1$ [$\,$Scenario I$\,$]. Short-dashed
line is $J-\langle \rho \rangle$ relationship for uniform hopping-rate systems,
Eq.(\protect{\ref{eq:j0}}). 
} 
\label{fig:jvsrho006}
\end{figure}

Furthermore, data within the region of validity of Scenario II can be compared with
the predictions of Eqs.~(\ref{eq:jvsth2}) and~(\ref{eq:rhovsth2}). We used
$\theta=0.06$. One sees in Fig.~\ref{fig:jvsrho006} 
that the agreement between theory and
extrapolated numerical results is indeed excellent. The prediction
of a plateau for Scenario I is also borne out by numerics, within error bars.
One cannot see unequivocal evidence here for a reentrant behavior similar
to that found in Fig.~\ref{fig:jthext}. It is possible that such an effect,
if present, is smaller than for the cases depicted in the latter Figure. This
would be in line with the observation that the amplitude of the reentrance
decreases with decreasing $\theta$~.

Fig.~\ref{fig:jvsrho006} also shows the
$J - \langle\rho\rangle$ relation for uniform hopping-rate systems, for comparison. 

A current-density diagram very similar to Fig.~\ref{fig:jvsrho006} was obtained
in Ref.~\onlinecite{lak06}, for the partially asymmetric exclusion problem
with spatially-varying hopping rates. 

\section
{Open boundary conditions}
\label{sec:obc}

\subsection{Introduction}
\label{obcintro}

With open boundary conditions, the following additional
quantities are introduced: the injection (attempt) rate $\alpha$ at the left end,
and the ejection rate $\beta$ at the right one. 
Calling $\rho_L$, $\rho_R$ the stationary densities respectively at the 
left and right ends of the chain, one has for the current $J$ at the
boundaries, and anywhere inside:
\begin{equation}
\alpha\,(1-\rho_L)=J=\beta\,\rho_R\ .
\label{eq:obc}
\end{equation}
Scenarios I and II, regarding the  existence and location of an "envelope",
discussed in the preceding section, still apply here, with similar consequences
upon the system-wide current. The overall picture turns out to be rather like 
that for open systems with uniform hopping 
rate~\cite{derr98,sch00,derr93,ds04,rbs01,be07,nas02,ess05}:
a maximal-current phase arises for suitably large $\alpha$,
$\beta$ (where Scenario I takes hold); elsewhere, one has less-than-maximal
current, although with either low or high density, the latter
two subphases being separated by a coexistence line; Scenario II applies. 
See Fig.~\ref{fig:obcpd} and corresponding insets.

The robustness of the three-phase structure in the present case is in line with
the results of Ref.~\onlinecite{lak06}. In their study of the partially
asymmetric exclusion problem, with spatially-varying right- and left-
hopping rates $p(x)$ and  $q(x)$ respectively, those authors always found
three phases, as long as $p(x)-q(x)$ did not change sign.    

In the maximal current phase, the following specific features are noteworthy:
\par\noindent
(i)\ the steady-state system-wide density $\langle \rho \rangle$ is very close
to the value which, for PBC, corresponds to the lower limit of
validity of Scenario I. This is because, from the conditions given in
Eq.~(\ref{eq:obc}), for large
$\alpha, \beta$ one must have $\rho_L$ "large" and $\rho_R$ "small".
Thus the density profile essentially follows the lower branch of the
envelope function. Slight departures from that occur within short ("healing")
distances from the extremes, in order to comply with the exact values
dictated by Eq.~(\ref{eq:obc}). The latter effects account for the fact
that $\langle \rho \rangle$ is not strictly constant throughout the maximal-current
phase. Although Eq.~(\ref{eq:obc}) imposes the same constraints for systems
with uniform hopping rates, there the envelope is trivially $x-$independent, 
and $\langle \rho \rangle$ is close to $1/2$~\cite{sch00,mukamel,rbs01}.
\par\noindent
(ii)\ In contrast to Scenario I with PBC, the steady-state 
profiles here do not show a kink inside the system.
\par\noindent
(iii)\ Similarly to Scenario I with PBC, the $\tan$- like segment of the 
profile at the extreme left of the system is essential in the local density adjustment
near that edge. However, as just mentioned, such adjustment is here
imposed by Eq.~(\ref{eq:obc}), as opposed to the former case where the constraint
arises from demanding continuity of $\rho$ to obey PBC 
(combined with the existence of a kink further to the right).

\subsection{Theory and Scenarios}
\label{obctheo}
  
\begin{figure}
{\centering \resizebox*{2.6in}{!}{\includegraphics*{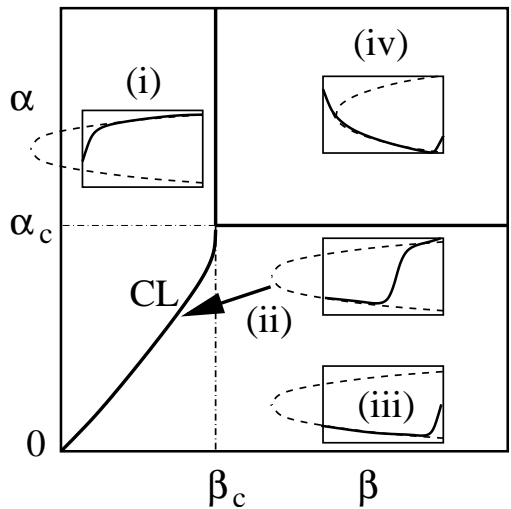}}}
\caption{
Schematic phase diagram for TASEP with hopping-rate gradient for
open boundary conditions.
Locations of phase boundaries are $\theta$- dependent [$\,$see
Eqs.~(\protect{\ref{eq:cl}})--(\protect{\ref{eq:a_cb_c}})$\,$]. 
CL stands for coexistence line (between high- and low- density phases).
The insets show typical density profiles for each phase (see text).
} 
\label{fig:obcpd}
\end{figure}

In the "low current" Scenario II, with the apex left of the system's 
left boundary ($x_0=-\lambda\,L/2$,
with $\lambda >1$), one has only $\tanh$ type solutions for all $x$;
$\sigma=2(\langle \rho \rangle-\frac{1}{2})$ is limited by the envelope $\pm k(x)$:
\begin{equation}
k(x)=\left[1 -\frac{4J}{p(x)}\right]^{1/2}=\left[\frac{(2\theta/L)(x-x_0)}
{1+(2\theta/L)\,x}\right]^{1/2}\ .
\label{eq:envel}
\end{equation}
So,
\begin{equation}
J=\frac{p(x)}{4}\,(1-k^2(x))=\frac{1}{8}\left(1+\frac{2\theta}{L}\,x\right)
\,(1-k^2(x))\ .
\label{eq:curr}
\end{equation}
Taking $x=x_0$, where $k(x_0)=0$, gives
\begin{equation}
J=\frac{1}{8}\,(1-\lambda\,\theta)\ ,
\label{eq:curr2}
\end{equation}
while evaluating Eq.~(\ref{eq:envel}) at $x=\mp L/2$ gives
\begin{equation}
k_{L,R}=k(\mp\frac{L}{2})=\left[\frac{\theta(\lambda\mp 1)}{1\mp \theta}\right]^{1/2}
\ ,
\label{eq:kpm}
\end{equation} 
which sets the upper ($\rho_{L,R}^>$) and lower ($\rho_{L,R}^<$) bounds 
for the density at the extremes:
\begin{equation}
\rho_{L,R}^{>,<}=\frac{1}{2}\left(1 \pm k_{L,R}\right)\ .
\label{eq:rhopm}
\end{equation}
With $\sigma(x)=k(x)\,\tanh (K(x)-K(a))$, where $x=a$ is the position of the kink,
and $K= \int k(x)\,dx$, one can see from the insets in Fig.~\ref{fig:obcpd}
 that, considering the
situations corresponding to profiles types (i), (ii), and (iii) shown there, 
the following constraints hold:
\begin{eqnarray}
\begin{cases}{
\rho_L^< \leq \rho_L \leq \rho_L^>\qquad\quad\ {\rm (i)}} \cr 
{\rho_L = \rho_L^<\qquad\quad\qquad {\rm (ii),(iii)}}
\end{cases}\ ; \nonumber \\ \noindent
\begin{cases}{
\rho_R = \rho_R^>\qquad\qquad\quad {\rm (i),(ii)}} \cr
{\rho_R^< \leq \rho_R \leq \rho_R^>\qquad\quad {\rm (iii)}} 
\end{cases}\ . 
\label{eq:constraints}
\end{eqnarray}
So, using Eqs.~(\ref{eq:obc}) and~(\ref{eq:curr}) we obtain, for the possible
values of  $\alpha$ and $\beta$ in the three situations:
\begin{eqnarray}
\alpha_{\rm (i)} \in \left[\frac{J}{1-\rho_L^<}\,,\,\frac{J}{1-\rho_L^>}\right] =
\frac{1}{4}(1-\theta)\,\left[1-k_L ,1 +k_L\right]\ ;\nonumber
\\
\alpha_{\rm (ii),(iii)} =\frac{J}{1-\rho_L^<} = \frac{1}{4}(1-\theta)\,(1-k_L)\ ;
\nonumber \\
\beta_{\rm (i),(ii)} =\frac{J}{\rho_R^>} = \frac{1}{4}(1+\theta)\,(1-k_R)\ ;
\nonumber \\
\beta_{\rm (iii)} \in \left[\frac{J}{\rho_R^>}\,,\,\frac{J}{\rho_R^<}\right] =\frac{1}{4}
(1+\theta)\,\left[1-k_R ,1 +k_R\right]\ .\nonumber
\\
\label{eq:alpha_beta}
\end{eqnarray}
For the coexistence line (CL), in which the kink lies wholly inside the system, 
i.e., profile type (ii) above, the results established in Eqs.~(\ref{eq:alpha_beta}),
together with Eqs.~(\ref{eq:curr2}) and~(\ref{eq:kpm}), give the current:
\begin{equation}
J_{\rm CL}= \alpha\left(1-\frac{2\alpha}{1-\theta}\right)
=\beta\left(1-\frac{2\beta}{1+\theta}\right);
\label{eq:jcl}
\end{equation}
and the equation for the CL shape as follows:
\begin{equation}
 \frac{2\alpha^2}{1-\theta}-\alpha=\frac{2\beta^2}{1+\theta}-\beta\ .
\label{eq:cl}
\end{equation}
In this "low current" Scenario II, $\lambda >1$ and, from Eq.~(\ref{eq:curr2}),
$\lambda \leq 1/\theta$. So the extent of the CL in $(\alpha,\beta)$
parameter space, and the current there, are limited to:
\begin{equation}
0 \leq \alpha\,\left(1 - \frac{2\alpha}{1-\theta}\right)= 
\beta\,\left(1 - \frac{2\beta}{1+\theta}\right) = J_{\rm CL} \leq \frac{1}{8}
(1-\theta)\ .
\label{eq:CLbounds}
\end{equation}
The same form of current, Eq.~(\ref{eq:curr2}), and the same limitation
$1 \leq \lambda \leq 1/\theta$, apply for the high density and low density
sub-phases (corresponding to profiles of types (i), (iii)) which the
coexistence line separates in this low-current Scenario II. Actually $\lambda=1$
is the boundary between  maximal (plateau) current phase (corresponding
to profiles of type (iv)) and the lower 
current phase(s). Since for $\lambda=1$, $k_L=0$, 
$k_R=\left(2\theta/(1+\theta)\right)^{1/2}$, using Eq.~(\ref{eq:alpha_beta}) the
phase boundaries are (in addition to the coexistence line):
\begin{equation}
\alpha=\alpha_c(\theta)\ ,\ \beta \geq \beta_c(\theta)
\label{eq:phb1}
\end{equation}
(between subphase (iii) and maximal current phase), and
\begin{equation}
\beta=\beta_c(\theta)\ ,\ \alpha \geq \alpha_c(\theta)
\label{eq:phb2}
\end{equation}
(between subphase (i) and maximal current phase), where
\begin{equation}
\alpha_c= \frac{1}{4}\,(1-\theta)\ ,\quad \beta_c= \frac{1}{4}\,(1+\theta)\,
\left[1 -\left(\frac{2\theta}{1+\theta}\right)^{\frac{1}{2}}\right]\ .
\label{eq:a_cb_c}
\end{equation} 
From Eqs.~(\ref{eq:cl}) and~(\ref{eq:a_cb_c}), the slope of the CL is
unity at the origin, i.e. the same there as that for the uniform-rate case,
and diverges at the endpoint ($\alpha_c$, $\beta_c$).
\vskip 0.2truecm
Given $\alpha$ and $\beta$, 
Eqs.~(\ref{eq:curr2}), (\ref{eq:kpm}), and~(\ref{eq:alpha_beta})
give $J$ and $\lambda$, and then Eqs.~(\ref{eq:rhovsKR2}) and~(\ref{eq:rhovsth2}) 
can be used for the determination of  $\langle \rho \rangle$,
anywhere on the phase diagram where Scenario II applies.
For points on the CL, however, an adaptation is needed in order to account for
the presence of a kink inside the system. Then,
the amended form of Eq.~(\ref{eq:rhovsKR2}) reads:
\begin{equation}
2(\langle\rho \rangle-\frac{1}{2}) \approx 
\frac{1}{L} \{\,{\widetilde K} (\frac{L}{2}(\lambda+1))+
{\widetilde K} (\frac{L}{2}(\lambda-1))-2\,{\widetilde K}(A)\,\}\ ,
\label{eq:rhovsKR2b}
\end{equation}
where $X=A$ is the position of the kink. This can be found by keeping
track of the leading finite-size corrections (from the asymptotic values 
$\pm 1$) to the $\tanh$ 
forms at the ends [$\,$to obey the constraints given by Eq~(\ref{eq:obc})$\,$].
One gets the prediction $\langle \rho \rangle =\frac{1}{2}$ everywhere on the CL,
for any $\theta$.

\subsection{Numerical results}
\label{obcnr}

In numerical work with open boundary conditions, we kept to $\theta=0.2$.

Initially we investigated the shape of steady-state profiles deep inside
the high-density, low-density, and maximal current regions given in 
Fig.~\ref{fig:obcpd}, as well as at a point on the CL at $(\alpha,\beta)=
(0.1, 0.087868)$ [$\,$about halfway between the origin and the endpoint of the 
CL, see Eqs.~(\ref{eq:cl}) and~(\ref{eq:a_cb_c})$\,$].
We found profiles which conform respectively to  types (i), (iii), (iv), 
and (ii) shown in the Figure, in agreement with the theoretical results given
above.

Deep inside the maximal-current phase, at $\alpha=\beta=0.375$,
we carried out a finite-size scaling
analysis of steady state currents, using systems with $L \leq 8192$. The
finite-$L$ values $J_L$ thus obtained were very close to those corresponding
to PBC and $\theta=0.2$, for $\langle\rho\rangle$ in Scenario I. 
They approach the same extrapolated value $J_\infty =\frac{1}{8}(1-\theta)$
found there, with the same type of finite-size corrections, i.e. $J_L-
J_\infty \sim L^{-\psi}$, $\psi \approx 0.5$.  

Elsewhere on the phase diagram, we  calculated steady-state currents and densities
at selected points, using only $L=1024$. Results are shown in Table~\ref{t1}.

For the point on the CL, the central value of $\langle \rho \rangle$ is 
close to $1/2$, as predicted in Subsection~\ref{obctheo}, but 
the density fluctuations, associated to phase coexistence, are
apparently very large. Related profiles 
(at the relatively small value $\theta=0.2$ being used)
are consistent with the kink being 
located with roughly equal probability anywhere in the system, 
as in the $\theta=0$ case~\cite{be07,dqrbs08}.
An approximate calculation, 
assuming this~\footnote{It is the number of particles present in the system,
rather than the kink location, which is expected to have uniform distribution,
so our procedure is not expected to apply at large $\theta$.}, 
and replacing the envelope 
by one with $k_L$ and $k_R$ both set equal to their root-mean-square value, 
provides an estimate 
for the root-mean-square density deviation, 
$ \langle \delta \rho \rangle_{\rm rms}=0.17$, in line with the result quoted
in Table~\ref{t1}.
The relationship between the current $J_{CL}$, $\alpha$, and $\beta$ given in 
Eq.~(\ref{eq:jcl}) is verified to very good accuracy.

In the high-(HD)  and low-density (LD) phases, agreement between theory and numerics 
is excellent, in part because finite-size effects are small there,
where Scenario II holds.

For the set of three points inside the maximal current (MC) phase, the currents
are indeed close to each other, their value differing from the infinite-system
one $J_\infty=\frac{1}{8}(1-\theta)$ by well-understood finite-size corrections.
The corresponding densities are also very close, and in good accord with
the prediction that the corresponding profiles should essentially coincide 
with the lower branch of the envelope function. Recall that, for PBC,
this is expected to happen, at $\theta=0.2$, for $\langle \rho \rangle$ 
close to $0.3$ (see Section~\ref{sec:pbc}). The larger spread between densities in the 
MC phase, when compared to that between currents,  is to be expected (see comments 
in Subsection~\ref{obcintro}). One gets a smaller difference between numerical results
and theoretical predictions by looking at a point on the borderline between LD and 
MC phases (LD/MC). Even then, the agreement is not as close as that found
deep inside the HD and LD phases. Such effects reflect the $L^{-1/2}$ corrections
pertaining to Scenario I.

\begin{table}
\caption{\label{t1}
Average steady-state currents $J$ and densities $\langle \rho\rangle$ for systems
with $\theta=0.2$, $L=1024$, and assorted injection/ejection rates  $(\alpha,\beta)$.
$J_{\rm th}$ and $\langle \rho \rangle_{\rm th}$ refer, respectively, to
currents and average densities calculated by the theory given in 
Subsec.~\protect{\ref{obctheo}}.
Phases specified in column 1 are,  respectively: CL: coexistence line; HD: 
high-density; LD: low-density; MC: maximal-current. Refer to
Fig.~\protect{\ref{fig:obcpd}} and text.
}
\vskip 0.2cm
\begin{ruledtabular}
\begin{tabular}{@{}lcccccc}
Type & $\alpha$ & $\beta$  & $J$ & $J_{\rm th}$ & $\langle \rho\,\rangle$ &
 $\langle \rho\,\rangle_{\rm th}$ \\
CL &  $0.100$ & $0.087868$ & $0.0750(3)$& $\frac{3}{40}$ & $0.51(10)$ & $\frac{1}{2}$ \\
HD &  $0.400$ & $0.100$ & $0.0834(4)$ & $\frac{1}{12}$ & $0.781(1)$ & $0.78241 \dots$ \\
LD &  $0.100$ & $0.400$ & $0.0751(3)$ & $\frac{3}{40}$ & $0.188(1)$ & $0.18839 \dots$ \\
LD/MC & $0.200$ & $0.400$ & $0.1015(2)$ & $\frac{1}{10}$ & $0.297(1)$& $0.29245 \dots$ \\
MC &  $0.450$ & $0.450$ & $0.1036(1)$ & $\frac{1}{10}$ & $0.3113(7)$& $0.29245 \dots$ \\
MC &  $0.650$ & $0.250$ & $0.1036(1)$ & $\frac{1}{10}$ & $0.3120(6)$& $0.29245 \dots$ \\
MC &  $0.250$ & $0.650$ & $0.1031(1)$ & $\frac{1}{10}$ & $0.3074(7)$& $0.29245 \dots$ \\
\end{tabular}
\end{ruledtabular}
\end{table}   

\section{Discussion and Conclusions} 
\label{sec:conc}
We have developed a mean-field/adiabatic theory for the one-dimensional TASEP
with smoothly-varying hopping rates. Its application to the uniform-gradient
case is shown, upon comparison with  extrapolations to the $L \to \infty$ 
limit of numerical simulation data, to give very accurate 
results. Evidence for this is exhibited especially in Figs.~\ref{fig:varprof02},
\ref{fig:rho05ext}, \ref{fig:jthext}, \ref{fig:jvsrho006}, and Table~\ref{t1}. 
Thus, for PBC
it appears that the $J-\langle\rho\rangle-\theta$ relationship given
by Eqs.~(\ref{eq:jvsth2}) and~(\ref{eq:rhovsth2}) is exact for Scenario II
of a $\langle\rho\rangle$- dependent current. While simulations essentially
find the constant-current plateau predicted for Scenario I with PBC
(at values of $J$ in full accord with theory), a small amount of
nonmonotonic dependence of $J$ on $\langle \rho \rangle$, near the edge of the
corresponding region, appears to be present. Although extrapolation of finite-system
current results turns out to be plagued with convergence issues precisely in this
region, a systematic trend is found towards increasing values of the calculated
overshoot as $\theta$ increases (see Fig.~\ref{fig:jthext}). Thus one cannot
definitely discard the possibility that such overhangs are real effects. 

Being mean field in character, the theory presented here
cannot predict, e.g., current fluctuations~\cite{dl98,ess11}, nor
fluctuation-related finite-size corrections. However, our numerical evidence
shows that in the plateau region, i.e., within  Scenario I 
(both for PBC and open BC's), the dominant finite-size
current corrections are of order $L^{-\psi}$, $\psi \approx 0.5$. This indicates
that an additional mechanism is present, whose effects obscure the usual
(uniform- hopping rate) fluctuation-induced $L^{-1}$  terms 
[$\,$the latter are clearly identified
in our numerics, not only for $\theta=0$,  but also wherever Scenario II holds$\,$].

The mean-field theory explains the $L^{-1/2}$ corrections as 
arising from a $\tan-$like part of the profile
which lies inside the system only in Scenario I. This occurs near the envelope
apex, in a region of
width $\pi/{\widetilde q}(X)$, where ${\widetilde q}=d{\widetilde Q/dX}$, 
see Eqs.~(\ref{eq:kx_qx}) and~(\ref{eq:kx_qx2}). 
There, $X={\cal O}(1)$, $c={\cal O}(L)$, hence 
from Eqs.~(\ref{eq:lin_mu}), (\ref{eq:lin_gamma}), and~(\ref{eq:x0def}),
$\pi/{\widetilde q}(X) ={\cal O}(L^{1/2})$. 
One must quantify the subdominant size-dependent effects originating from this
region.

For PBC, notice that in the range of $x$ where the $\tan$-- like profile holds, 
$\sigma={\cal O}(1)$ so it gives a contribution to the integral 
for $L\,\langle \rho \rangle$ of order $L^{1/2}$, out of a total of order $L$. 
This provides a correction of relative size
$L^{-1/2}$ in $\langle\rho\rangle$ for given $J$. By inverting the 
$J-\langle\rho\rangle$ relationship (since for PBC it is the density which is fixed), 
one is left with the observed current corrections ${\cal O}(L^{-1/2})$.

The argument for open boundary conditions is slightly different, because 
$\langle \rho  \rangle$ is not fixed by initial conditions
and $J$ is determined by the boundary injection/ejection rates. So,  we
look directly at the current and its relationship with $\alpha$ and $\rho_L$, 
as given in Eq.~(\ref{eq:obc}),
since the $\tan$ solution applies near the left edge.
By Eq.~(\ref{eq:kx_qx2}), this is $\sigma \sim -q_w\,\tan\ q_w(x-{\rm const.})$,
where $q_w$ is the value of $q=(-X/(X+c))^{1/2}$ at $X=x-x_0={\cal O}(1)$, with 
$X <0$. So $q_w={\cal O}(L^{-1/2})$, again by Eqs.~(\ref{eq:lin_mu}), 
(\ref{eq:lin_gamma}), and~(\ref{eq:x0def}). To provide the required injection
current, one must have $\sigma={\cal O}(1)$ near the left edge, while only a small
change $\Delta x$ in position, of order 
$\Delta x \approx 1/q_w ={\cal O}(L^{1/2})$ will take the $\tan \to 0$. The 
upshot is that the average change in $\sigma_L$ caused by a change of ${\cal O}(1)$
in $x_0$ is $\sim L^{-1/2}$. Hence with open boundary conditions, 
whenever Scenario I applies, the finite-size
correction in $J=\frac{\alpha}{2}\,(1-\sigma_L)$ is ${\cal O}(L^{-1/2})$~.

By similar arguments one finds that, when a kink is present, its
width generally gives corrections of order $L^{-1}$ to $\langle\rho\rangle$.
On the other hand, corrections coming from the region where the validity of 
the adiabatic 
approximation breaks down are of order $L^{-2/3}$. Since these only occur
when the apex is inside the system, i.e., when the $L^{-1/2}$ $\tan$- originated
terms are present as well, they are dominated by the latter.   

In closing, we note that a number of extensions of this study suggest themselves.
Steady state behavior for other spatial dependences of rates, particularly wells,
should be amenable to similar procedures. The same is true for studies of
the dynamics. To develop the theory beyond the mean field limit is a more
formidable challenge, but for slowly varying rates the adiabatic approach should still
apply, possibly combined with existing exact methods for uniform systems.
Phenomenological domain-wall approaches~\cite{ksks98,ps99} would be a likely
way forward.

\begin{acknowledgments}
The authors thank R. R. dos Santos and Fabian Essler for helpful discussions.
S.L.A.d.Q. thanks the Rudolf Peierls Centre for Theoretical Physics,
Oxford, where most of this work was carried out, for the hospitality,
and CAPES for funding his visit. The research of S.L.A.d.Q. is financed 
by the Brazilian agencies CAPES (Grant No. 0940-10-0),  
CNPq  (Grant No. 302924/2009-4), and FAPERJ (Grant No. E-26/101.572/2010).
R.B.S. acknowledges partial support from EPSRC Oxford Condensed Matter
Theory Programme Grant EP/D050952/1. 
\end{acknowledgments}

\end{document}